\documentclass[twocolumn,showpacs,preprintnumbers,amsmath,amssymb]{revtex4}

\usepackage{graphicx}
\usepackage{dcolumn}
\usepackage{bm}

\begin{document}

\preprint{APS}

\title{Time-dependent Currents of a Single-electron Transistor in Dissipative Environments}
\author{Jung Hyun Oh}
\email{jungoh@iquips.uos.ac.kr}
\author{D. Ahn }
\author{S. W. Hwang}

\affiliation{Institute of Quantum Information Processing and
Systems, University of Seoul, 90 Jeonnong, Tongdaemoon-ku, Seoul
130-743, Korea}%

\date{\today}

\begin{abstract}
Currents of the single-electron transistors driven
by time-dependent fields via external dissipative circuits are investigated theoretically.
By expressing the external circuit in terms of driven harmonic oscillators 
and using the reduced-density operator method, we derive time- and environment-dependent tunneling rates
in the regime of sequential tunneling and present  expressions for 
both displacement and tunneling currents with these tunneling rates.
It is found that the dissipative environments affect tunneling currents in two ways; the determination 
of driving voltages at tunneling junctions and the depletion of particle-hole distribution functions.
Considering a simple dissipative circuit, we discuss the effects of the
environment on tunneling currents in both static and time-dependent cases.
\end{abstract}


\maketitle

\section{Introduction}

There have been considerable interests in a single-electron transistor 
because of its potential applications.
Since Coulomb blockade peaks in its conductance oscillation are very sensitive to 
a fraction of charges, it is expected to be one of promising candidates for a detector
in measuring quantum states of quantum computation and information processing\cite{shnirman}.
Recently, Schoelkopf {\it et. al.}\cite{schoelkopf}
developed a high sensitive single-electron transistor.
By introducing a $LC$-resonant circuit connected to a single-electron transistor,
they drove the system in the radio-frequency regime to overcome $1/f$ noise
and obtain a high sensitity of detecting charges.

From the theoretical point of view, such a single-electron transistor is also 
very interesting because one should consider the influence of
dissipative environments($LC$-resonant circuit with a cable resistance)
on current-voltage characteristics as well as effects of time-dependent external perturbations.
In the presence of the dissipative environments,
tunneling rates of quasiparticles are strongly affected because an additional energy is needed 
to excite the environments as shown in the case of single- and multi-junctions\cite{grabert}.
On the other hand, time-dependent perturbations affect phase coherence of quasiparticles in time
and its effects manifest itself in photon-assisted tunneling;
quasiparticles are also able to tunnel by absorbing or emitting photons\cite{bruder,jauho}.
Thus, quasiparticles in the single-electron transistor have two different energy-exchange mechanisms
when they tunnel through junctions and then, one should take into account these two mechanisms 
in calculating the tunneling rates simultaneously.

In this work we investigate effects of the dissipative environments and the time-dependent perturbations
on current-voltage characteristics of a single-electron transistor by calculating the tunneling rate
and present expressions for currents at each electrode.
In calculating the tunneling rate we make two assumptions for our model of the single-electron transistor.
Firstly, tunneling barriers are considered so opaque that quasiparticles in each electrode
are well localized there and their motions can be described by separated Hamiltonians.
In other word, the tunneling barriers have a resistance much larger than the resistance quantum $R_K=h/e^2$.
Secondly, it is assumed that time between successive tunneling events is much larger than
the charge relaxation time of the dissipative environments.
This assumption makes the problem easy to
treat the dissipative environments as heat reservoirs being in thermal equilibrium. 
However, we still treat quasiparticles confined in the region of a quantum dot as in non-equilibrium.

The paper is organized as follows. 
We first describe the Hamiltonian of the single-electron transistor in Section II. 
By expressing the dissipative environments in terms of driven harmonic oscillators, we give the Hamiltonian
separating it into an interested system part and its environment,
and derive time- and environment-dependent tunneling rates using a reduced-density operator method.
In Section III, we present expressions for currents flowing in each electrode in terms of
displacement and tunneling components.
As an applilication of our expressions, tunneling currents driven by
a simple dissipative circuit are examined in Section IV, and then, a brief summary is given in Section V.

\section{Hamiltonian}

In Fig. 1, we show a typical drawing of the single electron transistor
driven by time-dependent voltages via possible dissipative elements.
The Hamiltonian of the entire system is modeled by 
${\cal H}={\cal H}_{qp}+{\cal H}_{RLC}(t)+{\cal H}_T$ where
the first two terms describe the motion of the system in the absence of tunnelings and
the last one is the tunneling Hamiltonian.
In the absence of tunneling, the system is viewed as just a simple electronic circuit(a lumped-circuit)
because all tunneling junctions are considered as capacitors.
Then, its motion is
described by two independent degrees of freedom;
microscopic and macroscopic variables\cite{ingold}.
\begin{figure}
\centering
\includegraphics[width=0.4\textwidth]{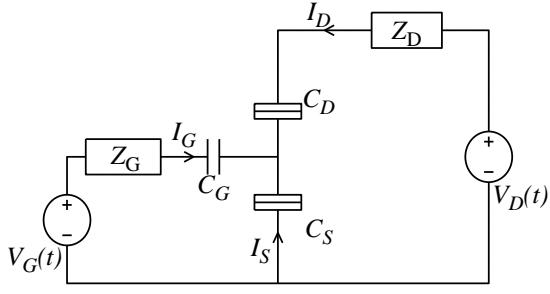}
\caption{\label{fig1}
A typical drawing of the single-electron transistor is shown where 
time-dependent voltages are applied to a quantum dot
via possible dissipative elements connected to drain and gate electrodes, respectively.
}
\end{figure}

By the microscopic variables, we mean those of quasiparticles in conductors of Fig. 1.
In our system, since quasiparticles relevant to tunneling are distributed
in a island called a quantum dot and adjacent electrodes (source and drain), 
we denote their Hamiltonian as ${\cal H}_{qp}={\cal H}_{dot}+{\cal H}_{lead}$ where 
${\cal H}_{dot}= \sum_l\epsilon_l d_l^\dagger d_l+V_{ee}(d_l^\dagger,d_l)$ and
${\cal H}_{lead}= \sum_{\alpha=S,D}\sum_k\epsilon_{k\alpha} a_{k\alpha}^\dagger a_{k\alpha}$.
Here, $d_l$ and $a_{k\alpha}$ ($d_l^\dagger$ and $a_{k\alpha}^\dagger$)
are the annihilation(creation) operators for quasiparticles
in the quantum dot with an energy $\epsilon_l$ and
in the electrodes $\alpha$ with an energy $\epsilon_{k\alpha}$, respectively.
In the quantum dot we denote the electron-electron interaction by 
$V_{ee}(d_l^\dagger,d_l)$ which in the most case is approximated by
a simple Coulomb-blockade model $Q^2/2C_\Sigma$ with excess charges $Q$ and $C_{\Sigma}=C_D+C_S+C_G$.
In the metallic electrodes, energies of quasiparticles are assumed to
be measured relative to their chemical potentials
and independent of time-dependent external perturbations by considering 
small frequencies not to excite plasmon.

On the other hand, macroscopic variables represent
charges on capacitors and flux through inductors in the lumped-circuit.
To describe their motion, we model the dissipative elements of $Z_D(\omega)$ and $Z_G(\omega)$ 
in Fig. 1 with different sets of simple harmonic oscillators ({\it i.e.} $LC$-circuits) 
following Caldeira and Leggett\cite{caldeira}.
Then, starting from the Lagrangian formulation where generalized coordinates are chosen to
be $\phi(t)=\int dt e v(t)/\hbar $ for a voltage $v(t)$ across each capacitor, 
the Hamiltonian ${\cal H}_{RLC}(t)$  of the macroscopic part is expressed as,
\begin{eqnarray}
{\cal H}_{RLC}(t) = \frac{Q_1^2}{2C_1}+\frac{Q_2^2}{2C_2}~~~~~~~~~~~~~~\nonumber\\
+\sum_{m=1}^M  \Big [\frac{q^{2}_m}{2c_m}+
\frac{1}{2L_m}\frac{\hbar^2}{e^2}\{\phi_1+\phi_2+\varphi_m-\psi_G(t)\}^2\Big] \nonumber \\
+\sum_{n=M+1}^{N+M} \Big[ \frac{q_n^2}{2c_n}+
\frac{1}{2L_n}\frac{\hbar^2}{e^2}\{\phi_1+\varphi_n-\psi_D(t)\}^2 \Big ]
\label{eq:HRLC}
\end{eqnarray}
where charges $Q_j$ and $q_n$ are conjugate to phases $\hbar\phi_j/e$
and $\hbar\varphi_j/e$, respectively, and satisfy the commutation relations of
$[\phi_j,Q_j]=ie$ and $[\varphi_n,q_n]=ie$.
The external perturbation $V_\alpha(t)$ are treated as a classical field and 
involved in externally driving phases $\psi_D(t)$ and $\psi_G(t)$ as,
\begin{eqnarray}
\psi_D(t) &=& \frac{e}{\hbar}\int_0^t d\tau V_D(\tau)\nonumber\\
\psi_G(t) &=& \frac{e}{\hbar} \frac{C_D+C_S}{C_D} \int_0^t d\tau V_G(\tau).
\end{eqnarray}
The first two terms in Eq. (\ref{eq:HRLC}) are additional charging energies on the capacitors
connected to each electrode
with effective capacitances $C_1=C_D C_S/(C_D+C_S)$ and $C_2=C_D^2 C_G/C_{\Sigma}(C_D+C_S)$,
whereas the remaining terms describe charging energies on capacitors with a capacitance $c_m$
and magnetic energies of inductors with a inductance $L_m$
in the dissipative elements, $Z_D(\omega)$ and $Z_G(\omega)$.
In fact, the charges $Q_1$, $Q_2$, and $Q$ in ${\cal H}_{RLC}$ describe
charges $Q_\alpha$ ($\alpha=S,D,G$) accumulated on a plate of the capacitor
connected to the electrode $\alpha$ in Fig. 1 and they are related to each other by,
\begin{eqnarray}
\left(
    \begin{array}{c}
Q_1\\
Q_2\\
Q  \\
     \end{array}
\right)
 = 
\left(
    \begin{array}{ccc}
 \frac{C_1}{C_D} & -\frac{C_1}{C_S} &  0        \\
-\frac{C_2}{C_D} & -\frac{C_2}{C_D} & \frac{C_D}{C_\Sigma}     \\
 -1      &  -1      & -1        \\
     \end{array}
\right)
\left(
    \begin{array}{c}
Q_D\\
Q_S\\
Q_G\\
     \end{array}
\right).
\label{eq:QandQ}
\end{eqnarray}
Then, classical relaxation equations for the charges $Q_\alpha(t)$ are exactly recovered from the
Heisenberg equation of motions. To do this, 
the impedances $Z_G(\omega)$ and $Z_D(\omega)$ at a angular frequency $\omega$ are anticipated to have
sets of parameters $\{c_n,L_n\}$ and $\{c_m,L_m\}$ like\cite{grabert},
\begin{eqnarray}
\frac{1}{Z_G^\prime(\omega)} &=& \frac{C_D^2}{(C_D+C_S)^2 Z_G(\omega)}
= \sum_{m=1}^M \frac{i\omega c_m}{1-c_m L_m \omega^2},\nonumber \\
\frac{1}{Z_D(\omega)} &=& \sum_{n=M+1}^{N+M} \frac{i\omega c_n}{1-c_n L_n \omega^2}.
\end{eqnarray}

The tunneling part of the Hamiltonian ${\cal H}_T$ may be given by,
\begin{eqnarray}
{\cal H}_T = \sum_{\alpha kl}  \Big [
T_{kl}^\alpha a_{k\alpha}^\dagger d_l e^{-i \phi_\alpha}+{\rm H.c.} \Big ]
\end{eqnarray}
where  $T_{kl}^\alpha$ denotes the element of tunneling matrix between a state 
$\mid k\rangle$ in the electrode $\alpha$
and a single particle state $\mid l\rangle$ in the quantum dot.
Here, the phase operator $\phi_\alpha$ $(\alpha=S,D)$ is defined in terms of 
$\phi_1$ and $\phi_2$ through
the relation,
\begin{eqnarray}
\left(
    \begin{array}{c}
\phi_D\\
\phi_S\\
     \end{array}
\right)
 = 
{\bf A }
\left(
    \begin{array}{c}
\phi_1\\
\phi_2\\
     \end{array}
\right)
=
\left(
    \begin{array}{ccc}
 \frac{C_1}{C_D} & -\frac{C_2}{C_D}\\
-\frac{C_1}{C_S} & -\frac{C_2}{C_D}\\
     \end{array}
\right)
\left(
    \begin{array}{c}
\phi_1\\
\phi_2\\
     \end{array}
\right).
\label{eq:dmatrix}
\end{eqnarray}
Thereby, the operator $e^{i\phi_\alpha}$ increases excess charges by the elementary charge $e$ 
in the electrode $\alpha$ for every tunneling event because they satisfy the commutation relation
$[\phi_\alpha,Q_\alpha]=ie$.

In reality, the form of the Hamiltonian ${\cal H}={\cal H}_{qp}+{\cal H}_{RLC}(t)+{\cal H}_T$ is not
adequate in obtaining expressions of currents
because the time-dependence of ${\cal H}_{RLC}(t)$ makes the problem complicated.
To circumvent this, we perform a standard time-dependent unitary transformation with,
\begin{eqnarray}
{\cal U}(t)&=& 
\prod_{j=1}^{2} \prod_{m=1}^{N+M} 
\exp\{i x_j(t)\phi_j+ ix_m^\prime(t)\varphi_m \} \nonumber\\
&&\exp\{-iu_j(t) Q_j/e-i u_m^\prime(t) q_m/e \}.
\end{eqnarray}
Here,  the sets of time-dependent functions $\{x_j(t),x_j^\prime(t)\}$ and $\{u_m(t),u_m^\prime(t)\}$
are chosen to rotate the Hamiltonian ${\cal H}_{RLC}(t)$ into the form irrelevant of time,
${\cal H}_{RLC}(0)$.
By a straightforward calculation for the rotated Hamiltonian ${\cal H}_R(t)$,
\begin{eqnarray}
{\cal H}_R(t) = {\cal U}^\dagger(t)
{\cal H} {\cal U}(t)-i\hbar {\cal U}^\dagger(t)\frac{\partial {\cal U}(t)}{\partial t}
\end{eqnarray}
one can show that a time-independent form of ${\cal H}_{RLC}(0)$ is 
obtained by choosing $u_j(t)$ just as the phase difference across 
capacitors forced by $\psi_D(t)$ and $\psi_G(t)$ in the absence of tunneling and
$x_j(t)=C_j \hbar \dot{u}_j/e^2 $.
That is, $u_j(t)$ is chosen by,
\begin{eqnarray}
\left(
\begin{array}{c}
u_1(t) \\
u_2(t) \\
\end{array}
\right)
=
{\bf A}^{-1}
\left(
\begin{array}{c}
y_D(t) \\
y_S(t) \\
\end{array}
\right),~~~~~~~~~~~\nonumber\\
y_\alpha(t) = \frac{e}{\hbar}\int^t_0 d\tau \Big\{
(\delta_{\alpha,D}-\frac{C_D}{C_\Sigma})V_D^0-\frac{C_G}{C_\Sigma}V_G^0 +g_\alpha(\tau)\Big\}
\end{eqnarray}
where we separate the external perturbation into  DC and AC parts, $i.e.$, $V_\alpha(t)=V_D^0+v_\alpha(t)$.
Here, $g_\alpha(\tau)$ describes voltage differences at tunneling junctions
exerted by the AC part of the voltage sources and is given in its Fourier component as,
\begin{eqnarray}
\left(
\begin{array}{cc}
 \frac{C_S+C_G}{C_\Sigma} &-\frac{C_D}{C_\Sigma} \\
-\frac{C_G}{C_\Sigma}     & -\frac{C_G}{C_\Sigma}\\
\end{array}
\right)
{\bf Z}^{-1}
\left(
\begin{array}{c}
\tilde{g}_D(\omega)\\
\tilde{g}_S(\omega)\\
\end{array}
\right)
=
\left(
\begin{array}{c}
\tilde{v}_D/Z_D \\
\tilde{v}_G/Z_G \\
\end{array}
\right)
\label{eq:galpha}
\end{eqnarray}
where ${\tilde v}_\alpha(\omega)$ is a Fourier component of the AC part of the external perturbation and
the matrix ${\bf Z}$ is defined as,
\begin{eqnarray}
{\bf Z} = 
{\bf A} \left(
\begin{array}{cc}
i \omega C_1+Z_D^{-1}+Z_G^{\prime-1} & Z_G^{\prime-1} \\
Z_G^{\prime-1} & i \omega C_2+Z_G^{\prime-1} \\
     \end{array}
\right)^{-1}{\bf A}^T.
\label{eq:ImZ}
\end{eqnarray}
Finally, under the unitary transformation ${\cal U}(t)$, the total Hamiltonian becomes,
\begin{eqnarray}
{\cal H}_R(t) &=&
 {\cal H}_{dot}+ {\cal H}_{lead}
+{\cal H}_{RLC}(0)+{\cal H}_{T}^\prime(t).
\label{HR}
\end{eqnarray}
Here, ${\cal H}_{T}^\prime(t)$ is the tunneling Hamiltonian rotated by the unitary transformation ${\cal U}$
and now involves effects of the time-dependence perturbations as,
\begin{eqnarray}
{\cal H}_{T}^\prime(t) &=&\sum_{\alpha=S,D}\sum_{kl}  \Big [
T_{kl}^\alpha(t)d_l a_{k\alpha}^\dagger e^{-i\phi_\alpha}+{\rm H.c.} \Big ]
\end{eqnarray}
with $T_{kl}^\alpha(t)= T_{kl}^\alpha e^{-iy_\alpha(t)}$.

\subsection{Ensemble average}

In order to evaluate an ensemble average of physical quantities, we use the 
reduced-density operator method\cite{ahn}.
In this method,  the entire system 
is divided into an interested system part being in non-equilibrium, a heat bath, and their interaction;
$ {\cal H}_{R}= {\cal H}_{S}+ {\cal H}_{B}+{\cal H}_{SB}$.
Then, the effective density operator describing the interested system is obtained by averaging
the Liouville equation over the heat bath.
Since the heat bath is considered as in thermal equilibrium, the ensemble average of physical 
quantities is expressed as the sum of expectation values between quantum states
weighted with the reduced-density matrix.

In our case, we consider the system part as quasiparticles in the quantum dot, 
${\cal H}_S={\cal H}_{dot}$, and the heat bath 
as the combination of quasiparticles in the electrodes
and the lumped-circuit, {\it i.e.} ${\cal H}_{B}={\cal H}_{lead}+{\cal H}_{RLC}(0)$.
Then, the system and the heat bath are coupled by the time-dependent tunneling 
Hamiltonian ${\cal H}_{SB}(t)={\cal H}_T^\prime(t)$.
By adopting very opaque tunneling junctions (weak coupling of ${\cal H}_{SB}$) such that
a typical tunneling period
is much larger than the charge relaxation time of the heat bath,
one can treat the heat bath as being in thermal equilibrium.
In this case, its density operator $\rho_B$ is proportional to $e^{-\beta {\cal H}_B}$ 
at an inverse temperature $\beta$.

The density operator describing the interested system $\rho(t)$ is obtained by
tracing the density operator $\rho_{tot}(t)$ for the entire system
over the heat bath, $\rho(t)={\rm tr}_B\{ \rho_{tot}(t)\}$,  and
its equation of motions is derived from the Liouville equation.
In a non-Markovian form,
$\rho(t)$  is given by\cite{ahn},
\begin{eqnarray}
\frac{d\rho(t)}{dt} = \frac{1}{i\hbar} [{\cal H}_S,\rho(t)] +{\cal C}(t).
\label{eq:req}
\end{eqnarray}
Here, the generalized scattering operator  ${\cal C}(t)$ stands for,
\begin{eqnarray}
{\cal C}(t) &=& \frac{1}{\hbar^2}\int_{-\infty}^t d\tau
{\rm tr}_B\Big \{ \Big [ {\cal H}_{SB}(t),[\rho^I(\tau,t)\rho_B, {\cal H}_{SB}^I(\tau,t) ] \Big ] \Big \}
\nonumber\\
&&+O({\cal H}_{SB}^4)
\label{eq:scatt}
\end{eqnarray}
where $\rho^I(t)$ and ${\cal H}^I_{SB}(t)$ are their interaction pictures
of $\rho(t)$ and ${\cal H}_{SB}(t)$, respectively;
\begin{eqnarray}
\rho^I(\tau,t) &=&  e^{-i {\cal H}_S(t-\tau)/\hbar}\rho(\tau)
e^{i {\cal H}_S(t-\tau)/\hbar},\\
{\cal H}_{SB}^I(\tau,t) &=& 
e^{-i({\cal H}_S+{\cal H}_{B})(t-\tau)/\hbar} {\cal H}_{SB}(\tau)
\nonumber\\
&&e^{i ({\cal H}_S+{\cal H}_{B})(t-\tau)/\hbar}.
\end{eqnarray}
Here, the forth order contributions of the interaction Hamiltonian ${\cal H}_{SB}(t)$ are ignored
by taking into account opaque tunneling junctions.
Then, with this density operator, one can express the ensemble average of an arbitrary operator $\cal{O}$
in terms of relevant system operators by replacing the part of the heat bath with their equilibrium values.
That is,
$\langle{\cal O}\rangle ={\rm tr}_S\{{\rm tr}_B \{ {\cal O} \rho_{tot}(t)\}\} =
{\rm tr}_S\{{\cal O}_S \rho(t)\}$ where
${\rm tr}_S$ means the average of the operators over the system and
${\cal O}_S$ is a relevant system operator

\subsection{Expectation values for heat-bath operators}

For the ensemble averages of physical quantities,
one need to evaluate several time-correlations between heat-bath operators.
First, for those of the $RLC$-circuit, the correlation between phase operators of
${\rm tr}_B\{  e^{-i\phi_\alpha} e^{i\phi_\alpha(t)} \rho_B\}$
is necessary to calculate its effect on tunneling.
Here, $\phi_\alpha(t)$ is the Heisenberg operator of $\phi_\alpha$ with respect to ${\cal H}_{RLC}$.
Since the $RLC$-circuit of Eq. (\ref{eq:HRLC}) is considered as the sum of independent harmonic
oscillators in equilibrium,
this correlation function can be rewritten as\cite{ingold}
\begin{eqnarray}
{\rm tr}_B\{  e^{-i\phi_\alpha} e^{i\phi_\alpha(t)}\rho_B \}
= e^{{\rm tr}_B\{(\phi_\alpha\phi_\alpha(t)-\phi_\alpha\phi_\alpha)\rho_B \} }.
\end{eqnarray}
Then, based on the linear response theory, one can show that
the fluctuation of $J_\alpha(t) \equiv {\rm tr}_B \{ (\phi_\alpha\phi_\alpha(t)-\phi_\alpha\phi_\alpha)\rho_{B}\}$
is directly related to the dissipation of the $RLC$-circuit(Fluctuation-dissipation theorem).
From response functions of the $RLC$-circuit of
Eq. (\ref{eq:HRLC}) together with Eqs. (\ref{eq:QandQ}) and (\ref{eq:dmatrix}), $J_\alpha(t)$ is given by,
\begin{eqnarray}
J_\alpha(t) = \int_{-\infty}^\infty \frac{d\omega}{\omega}
\frac{{\rm Re}Z_\alpha^t(\omega)}{R_K} \frac{e^{i\omega t}-1}{1-e^{-\hbar\omega\beta}}
\label{eq:jalpha}
\end{eqnarray}
where $Z_\alpha^t(\omega)$ is an effective impedance of the $RLC$-circuit
seen from the tunnel junction $\alpha$\cite{grabert}
and is equal to diagonal elements of the impedance matrix ${\bf Z}$ of Eq. (\ref{eq:ImZ});
$Z_D^t(\omega)=Z_{11}(\omega)$ and $Z_S^t(\omega)=Z_{22}(\omega)$, respectively.
In reality, $Z_\alpha^t(\omega)$ has a slightly different form from the impedance
seen from a tunneling junction $\alpha$ in Fig. 1.
This is because the region of the quantum dot in Fig. 1 is independent of the $RLC$-circuit and does not contribute
to the fluctuation in the absence of tunneling.

As for the electrodes,  the following particle and hole evolutions are necesarry to evaluate tunneling currents,
\begin{eqnarray}
{\rm tr}_B\{ a^\dagger_{k\alpha} a_{k\alpha}(t)\rho_B \} &=&
f_{FD}(\epsilon_{k\alpha}) e^{-\gamma_\alpha\mid t\mid -\frac{i}{\hbar}\epsilon_{k\alpha} t}
\nonumber \\
{\rm tr}_B\{ a_{k\alpha} a^\dagger_{k\alpha}(t)\rho_B \} &=&
\{1-f_{FD}(\epsilon_{k\alpha})\} e^{-\gamma_\alpha\mid t\mid +\frac{i}{\hbar}\epsilon_{k\alpha} t}
\label{eq:ave}
\end{eqnarray}
where $a_{k\alpha}(t)$ is the Heisenberg operator of
$a_{k\alpha}$ with respect to ${\cal H}_{lead}$ and
$f_{FD}(\epsilon)=1/(1+e^{\epsilon \beta})$ is the Fermi-Dirac distribution function.
Here, the exponential decay of $e^{-\gamma_\alpha\mid t\mid}$ represents effects of tunneling
on states in the electrodes.
In reality, the heat bath as well as the quantum dot in our system are affected by tunneling.
As a result, the evolutions of quasiparticles in the heat-bath are different from those
in the isolated one and are usually represented by a finite life time.
The exponential decay is inserted by hand to represent such a effect on states\cite{bruder}.
The parameter $\gamma_{\alpha }$ is assumed to be the bare tunneling rate at the electrode $\alpha$,
\begin{eqnarray}
\gamma_{\alpha }(\epsilon)=\frac{2\pi}{\hbar}\sum_k\mid T^\alpha_{lk}\mid^2
\delta(\epsilon-\epsilon_{k\alpha}),
\end{eqnarray}
which  is usually considered as a constant independently of an energy within a so-called wide-band limit;
$\gamma_{\alpha }(\epsilon)=\gamma_\alpha$.

\subsection{Time- and environment-dependent master equation}

Now, we evaluate the reduced-density operator of Eq. (\ref{eq:req}) in the basis representation.
A simplified form is obtained when one expands the reduced-density operator 
in terms of many-body eigenstates($\mid\! r\rangle$ and $\mid\! s\rangle$)
of the quantum dot, $i.e.$,
\begin{eqnarray}
\rho(t)=\sum_{rs} P_{rs}(t) \mid r\rangle \langle s\mid.
\end{eqnarray}
Substituting this into Eq. (\ref{eq:req}) and then projecting it on one of diagonal components,
the occupation probability $P_{ss}(t)$ at a state $\mid\! s\!\rangle$ is given as a balanced form,
\begin{eqnarray}
\frac{d P_{ss}(t) }{dt} &=&\sum_{r\alpha\xi=+,-}  \Big [
\int_{-\infty}^0 d\tau P_{rr}(t+\tau) \Gamma^{\alpha \xi}_{rs}(t,\tau)-\nonumber\\
&&\int_{-\infty}^0 d\tau  P_{ss}(t+\tau) \Gamma^{\alpha \xi}_{sr}(t,\tau)\Big ]
\label{eq:master}
\end{eqnarray}
where the first term describes the increasing rate of the probability density
by transitions from other states while the second term is a decay rate due to transitions to others.
In deriving this result we disregard the contribution of off-diagonal components because
their effects are the forth order of the interaction Hamiltonian, $O({\cal H}_{SB}^4)$
and thus, the result does not contain coherent evolutions between many-body states which may be
caused by external perturbations.

In Eq. (\ref{eq:master}) the memory kernels of $\Gamma^{\alpha\pm}_{rs}(t,\tau)$
describe quasiparticle tunneling 
into $(+)$ or from $(-)$ the quantum dot through the barrier $\alpha$ 
to result in the transition from a state $\mid r\rangle$  to another state $\mid s\rangle$.
The detailed forms of $\Gamma^{\alpha\pm}_{rs}(t,\tau)$ are given by,
\begin{eqnarray}
\Gamma^{\alpha \pm}_{rs}(t,\tau)=
\frac{{\rm Re}}{\pi\hbar}
\int_{-\infty}^\infty d\epsilon~ \Lambda^{\alpha\pm}_{rs}(\epsilon+E_{rs}^{\alpha\pm})~~~~~~~~~~~~~~~
\nonumber\\
\exp \Big \{\gamma_\alpha\tau-\frac{i\epsilon}{\hbar}\tau+
J_\alpha(\pm\tau)-\frac{ie}{\hbar}\int_{t+\tau}^t d t^\prime  g_\alpha(t^\prime) \Big  \}
\label{eq:Gamma}
\end{eqnarray}
Here, $\Lambda^{\alpha\pm}(E)$ represent tunneling rates of
quasiparticles into($+$) or from($-$) the quantum dot with an energy gain of $E$ when
there is neither dissipative elements nor alternating perturbations 
in addition to negligible collision-broadening. 
In this case, the tunneling rates $\Lambda^{\alpha\pm}_{rs}(E)$ are reduce to
the widely used formula\cite{others};
\begin{eqnarray}
\Lambda^{\alpha\pm}_{rs}(E^{\alpha\pm}_{rs}) =\gamma_{\alpha} S^\pm_{rs} \Big\{ \frac{1}{2}\mp\frac{1}{2}\pm 
f_{FD}(E^{\alpha\pm}_{rs}) \Big\}
\end{eqnarray}
where coefficients $S^\pm_{rs}$ represent the selection rules of tunneling,
\begin{eqnarray}
S^+_{rs} = \sum_l\mid\langle r\mid d_l \mid  s\rangle\mid^2,~~~~~
S^-_{rs} = \sum_l\mid\langle s\mid d_l \mid  r\rangle\mid^2
\end{eqnarray}
and
 $ E_{rs}^{\alpha\pm} = \pm(E_s-E_r)+(\delta_{\alpha,D}-\frac{C_D}{C_\Sigma})eV_D^0-\frac{C_G}{C_\Sigma}eV_G^0$
are energy gains at a tunneling event with $E_r$ and $E_s$, eigenenergies of states
$\mid r\rangle$ and $\mid s\rangle$, respectively.
On the other hand, the exponential part of the integrand in Eq. (\ref{eq:Gamma}) includes
the effects of the dissipative environment and the alternating perturbations.
As indicated by the Tien-Gorden theory\cite{tien}, the external source contributes
the imaginary part in an argument of the exponential function via $g_\alpha(t)$, and
thus affects phases of electronic 
states. Whereas, the term $J_\alpha(t)$ in general have both real and imaginary parts, and thus
gives rise to the damping of states as well as the change of phases.
Roles of $ J_\alpha(t)$ in  $\Gamma^{\alpha \pm}_{rs}(t,\tau)$ are more easily understood
by transforming into its Fourier components as,
\begin{eqnarray}
 e^{J_\alpha(t)} = \int_{-\infty}^\infty d\omega e^{ i\omega t} P_\alpha(\omega).
\end{eqnarray}
Substituting this relation into Eq. (\ref{eq:Gamma}),
one can see that tunneling occurs with a weight of $P_\alpha(\omega)$
in the range between $\omega$ and $\omega+d\omega$ and corresponding energy gains become
$\epsilon \pm \hbar\omega+E_{rs}^{\alpha\pm}$.
Since the sum of $P_\alpha(\omega)$ over all frequency range is equal to one, 
$P_\alpha(\omega)$ can be interpreted as the probability density to exchange the energy $\hbar\omega$
between the system and its environment\cite{ingold}.
By considering energy differences between tunneling events,
one can see that $P_\alpha(\omega)$ in the positive(negative) frequency 
represents the probability to emit(asorbe) photons to(from) the environment.

From Eq. (\ref{eq:Gamma}), it is noted that effects of the dissipative environments on tunneling are
two folds.
The first is  the determination of the voltage difference $g_\alpha(t)$ at each tunneling junction
through Eq. (\ref{eq:galpha}).
The other is the probability density $P_\alpha(\omega)$ which is determined by the characteristic impedance
$Z_\alpha(\omega)$ of the environments via Eq. (\ref{eq:jalpha}).

\section{expressions for currents}

Now, we calculate currents in the electrodes in Fig. 1 where
a positive current at each electrode is  defined to flow into the quantum dot.
The current flowing in each electrode $\alpha$ consists of two different contributions; 
tunneling currents of quasiparticles $I^t_\alpha(t)$ and time-variation of charges on capacitors
in the lumped-circuit $I^d_\alpha(t)$ called displacement currents, $i.e.$, 
\begin{eqnarray}
I_\alpha(t) = I^d_\alpha(t)+I^t_\alpha(t).
\label{curr0}
\end{eqnarray}
Here, each component is calculated by time-derivatives for the ensemble average of 
particle numbers and charges;
\begin{eqnarray}
I^t_\alpha(t) &=& e\frac{d}{dt}\langle {\cal N}_\alpha\rangle_0
= e\frac{d}{dt}\Big\langle \sum_k a^\dagger_{k\alpha}a_{k\alpha}\Big\rangle_0 \nonumber\\
I^d_\alpha(t) &=& \frac{d}{dt}\langle Q_\alpha \rangle_0
\label{curr02}
\end{eqnarray}
where $\langle\ldots\rangle_0$ means the average over the Hamiltonian of ${\cal H}$.
With the above current expressions, the currents flowing into the quantum dot
are conserved even for the system subject to the time-dependent perturbations,
as emphasized by B\"{u}ttiker in his recent work\cite{buttiker}.
This can be shown by calculating the time-derivative of $\langle Q\rangle$, which gives
$d\langle Q\rangle /dt= \langle[Q,{\cal H}]\rangle/i\hbar= I_S^t(t)+I_D^t(t)$ reflecting the fact that
the increase of charges in the quantum dot is enabled by tunneling processes.
Alternatively, since $d\langle Q\rangle /dt$ is the sum of all displacement currents out of
the quantum dot from Eq. (\ref{eq:QandQ}), we obtain 
the conservation of the currents, $\sum_\alpha I_\alpha(t) = \sum_\alpha\{I_\alpha^t(t)+I_\alpha^d(t)\}
=0$ with $I_G^t(t)=0$.

By solving the Heisenberg equations of motion,
$d\langle Q_\alpha \rangle /dt= \langle[Q_\alpha ,{\cal H}]\rangle/i\hbar$,
 we can express the displacement currents in terms of the contributions of tunneling currents
and external perturbations. The results are, in its Fourier components of ${\tilde I}_\alpha^d(\omega)$,
\begin{eqnarray}
\left(
\begin{array}{cc}
 \frac{1}{i\omega C_D}+Z_D & -\frac{1}{i\omega C_S} \\
 \frac{1}{i\omega C_G}+Z_G &  \frac{C_S+C_G}{i\omega C_S C_G}+Z_G \\
\end{array}
\right)
\left(
\begin{array}{c}
 {\tilde I}^d_D(\omega)\\
 {\tilde I}^d_S(\omega)\\
\end{array}
\right)=~~~~~~~~~~~~~~\nonumber\\
\left(
\begin{array}{c}
 {\tilde v}_D(\omega)\\
-{\tilde v}_G(\omega)\\
\end{array}
\right)
\!-\!
\left(
\begin{array}{cc}
 Z_D & 0 \\
 \frac{1+i\omega C_G Z_G}{i\omega C_G} & \frac{1+i\omega C_G Z_G}{i\omega C_G}\\ 
\end{array}
\right)
\left(
\begin{array}{c}
 {\tilde I}^t_D(\omega)\\
 {\tilde I}^t_S(\omega)\\
\end{array}
\right).
\label{eq:dcurrent}
\end{eqnarray}
where ${\tilde I}^t_\alpha (\omega)$ is a Fourier component of a tunneling current $ I_\alpha^t(t)$.
Here, the first term on the right-hand side is the contribution from alternating  perturbations
while the second terms are resulted from tunneling.
Alternatively, the above results can be expressed in the equivalent circuit as shown in Fig. 2
by modeling the tunneling contributions as current sources.
Then, once tunneling currents $I_\alpha^t(t)$ are known,
the total currents at each electrode are  determined by applying basic circuit rules to Fig. 2.
\begin{figure}
\centering
\includegraphics[width=0.4\textwidth]{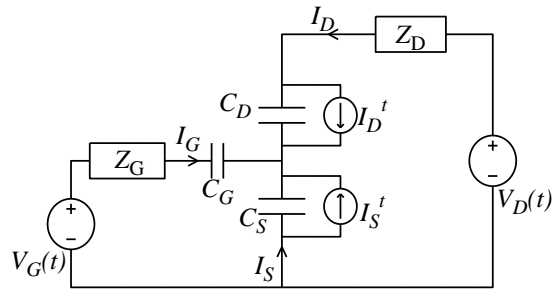}
\caption{\label{fig2}
For calculating both displacement and tunneling currents in the single-electron transistor of Fig. 1,
an equivalent circuit is shown, where components of tunneling currents are considered as current sources.
}
\end{figure}

Now we evaluate the tunneling component of the total current in Eq. (\ref{curr0}).
From the Heisenberg equation of motion, $d\langle e{\cal N}_\alpha \rangle_0 /dt= \langle[
 e{\cal N}_\alpha ,{\cal H}]\rangle_0/i\hbar$, the tunneling current $ I_\alpha^t(t)$ is calculated as,
\begin{eqnarray}
I_\alpha^t(t) &=& \frac{e}{i\hbar}\sum_{\alpha k l}\Big\langle 
T_{kl}^\alpha a_{k\alpha}^\dagger d_l e^{-i \phi_\alpha}-{\rm H.c.} \Big \rangle_0 \nonumber\\
 &=& \frac{e}{i\hbar}\sum_{\alpha k l}\Big\langle 
T_{kl}^\alpha(t) a_{k\alpha}^\dagger d_l e^{-i\phi_\alpha} -{\rm H.c.} \Big\rangle
\nonumber\\
&\equiv& \langle {\cal I}_\alpha(t) \rangle
\label{eq:tcurrent0}
\end{eqnarray}
where in the second line $\langle ...\rangle $ denotes an ensemble average over the rotated system ${\cal H}_R$.
By considering the second order contribution of ${\cal H}_{SB}$ or
only sequential tunneling processes, the expectation value of Eq. (\ref{eq:tcurrent0}) can be rewritten in
the reduced-density operator formalism as,
\begin{eqnarray}
I_\alpha^t(t) = \frac{1}{i\hbar}\int_{-\infty}^t \!d\tau \!{\rm tr}_S{\rm tr}_B\Big\{ 
{\cal I}_\alpha(t) [{\cal H}_{SB}^I(\tau,t),\!\rho_B\rho^I(\tau,t) ]
\Big\}.
\end{eqnarray}
Here, comparing this with Eq. (\ref{eq:tcurrent0}),  one can see that the part of the commutator
is resulted from the evolution of the total density $\rho_{tot}(t)$.
By replacing heat-bath operators with their equilibrium values and substituting the
reduced-density operator in the basis representation,
finally we obtain the tunneling current at the electrode $\alpha$ as,
\begin{eqnarray}
I_\alpha^t(t) = e\sum_{rs} \int_{-\infty}^0 \!d\tau\! \Big \{ \Gamma^{\alpha -}_{rs}(t,\tau)\!-\!
\Gamma^{\alpha +}_{rs}(t,\tau) \Big \} P_{rr}(t\!+\!\tau).
\label{eq:tcurrent}
\end{eqnarray}
As in the case of the occupation probabilities, off-diagonal contributions are also ignored because
their effects are the forth order of ${\cal H}_{SB}$.

Equations (\ref{eq:master}), (\ref{eq:Gamma}), and  (\ref{eq:tcurrent})
are main results of our work.
Based on the master equations of Eq. (\ref{eq:master}) and 
the memory kernel of Eq. (\ref{eq:Gamma}),
the displacement and tunneling currents flowing in each electrode of Fig. 1 
can be calculated using Eqs. (\ref{eq:dcurrent}) and (\ref{eq:tcurrent}), respectively.

\subsection{Expressions in the Fourier space}
For numerical calculations,
it is useful to introduce the Fourier transform of the memory kernel, Eq. (\ref{eq:Gamma}).
When external perturbations are periodic functions with an angular frequency $\omega_A$, we define
Fourier components as,
\begin{eqnarray}
{\tilde \Gamma}^{\alpha \pm}_{rs} (l,m) = \frac{\omega_A}{2\pi}
\int_0^{2\pi/\omega_A} dt
\int_{-\infty}^0 d\tau
~~~~~~~~~~~~~ \nonumber\\
\exp \big \{-i l\omega_A t+i m\omega_A \tau \big \}
{\tilde \Gamma}^{\alpha \pm}_{rs} (t,\tau).
\label{eq:Gamma0}
\end{eqnarray}
Then, by writing  the external perturbations in their Fourier components,
\begin{eqnarray}
\exp \big\{ -\frac{ie}{\hbar}\int_0^t d t^\prime  g_\alpha(t^\prime) \big  \}
=\sum_{n=-\infty}^\infty e^{ i n \omega_A t}\eta_\alpha(n),
\end{eqnarray}
we obtain ${\tilde \Gamma}^{\alpha \pm}_{rs} (l,m)$ as,
\begin{eqnarray}
{\tilde \Gamma}^{\alpha \pm}_{rs} (l,m) &=& \gamma_\alpha ~S^\pm_{rs}~ f_{lm}^{\alpha\pm}( E_{rs}^{\alpha\pm}).
\label{eq:GammaF}
\end{eqnarray}
Here, effective quasiparticle$(+)$ and hole$(-)$ distributions $f_{lm}^{\alpha \pm} (E)$ are defined by,
\begin{eqnarray}
f_{lm}^{\alpha \pm} ( E ) =  \frac{\delta_{l,0}}{2} \pm  \sum_n
\int_{-\infty}^\infty d\omega P_\alpha(\omega)~~~~~~~~~~~~~\nonumber\\
\Big [
\frac{\eta_\alpha (n)\eta_\alpha^* (n-l)}{2\pi i}
\psi^0 (z)+\{l,m\rightarrow -l,-m  \}^*  \Big ]
\label{eq:GammaFD}
\end{eqnarray}
where
\begin{eqnarray}
z = \frac{1}{2}+\frac{\hbar\gamma_\alpha\beta}{2\pi}+\frac{\beta}{2\pi i}
\{ E-(m+n)\hbar \omega_A\pm\hbar \omega\}\nonumber
\end{eqnarray}
and $\psi^0(z)$ is a digamma function.
Without the dissipative elements and the alternating perturbations
in addition to negligible collision-broadening $\gamma_\alpha$,
$f^{\alpha\pm}_{lm}(E)$ are reduced to $\frac{1}{2}\{1\mp {\rm tanh}(\frac{\beta E}{2})\}$,
the Fermi-Dirac distributions for particles($+$) and holes($-$).
However, in general, $f^{\alpha\pm}_{lm}$ is deviated from the Fermi-Dirac form due to 
the dissipative environments(described by $P_\alpha$)
as well as the finite life time of quasiparticles (described by $\gamma_\alpha$).

In terms of ${\tilde \Gamma}^{\alpha \pm}_{rs} (l,m)$,
the occupation probabilities and tunneling currents are calculated as,
\begin{eqnarray}
il\omega_A {\tilde P}_{ss}(l) = \sum_{r\alpha m }\sum_{\xi=+,-}\Big \{
{\tilde P}_{rr}(m) {\tilde \Gamma}^{\alpha \xi}_{rs} (l-m,m)
\nonumber\\
-{\tilde P}_{ss}(m) {\tilde \Gamma}^{\alpha \xi}_{sr} (l-m,m)\Big \}
\end{eqnarray}
and
\begin{eqnarray}
{\tilde I}_{\alpha}^t (l) = e \sum_{r s m }
{\tilde P}_{rr}(m) 
\Big \{ {\tilde \Gamma}^{\alpha -}_{rs} (l\!-\!m,m)\!-\!{\tilde \Gamma}^{\alpha +}_{rs} (l\!-\!m,m)
\Big \}
\end{eqnarray}
where we also expand $P_{ss}(t)$ and $I_{\alpha}^t(t)$ in their Fourier series;
$P_{ss}(t)=\sum_l e^{i l \omega_A t} {\tilde P}_{ss}(l)$ and 
$I_{\alpha}^t(t)=\sum_l e^{i l \omega_A t} {\tilde I}_{\alpha}^t(l)$. 
In special case of $P_\alpha(\omega)=\delta(\omega)$, $i.e.$ without dissipative elements in the circuit,
we find that the above results together with the tunneling rates of  Eq. (\ref{eq:GammaF}) are similar to the formalism
developed by Bruder and Schoeller\cite{bruder}.

\subsection{Time-convolutionless form}

Evaluating the integral of Eq. (\ref{eq:Gamma}), we derive another form of the memory kernel,
\begin{eqnarray}
\Gamma^{\alpha \pm}_{rs}(t,\tau)=
\gamma_\alpha S_{rs}^\pm \Big [ \delta(\tau)\mp \frac{\rm
csch( \frac{\pi \tau}{\hbar\beta})}{\hbar\beta} 
{\rm Im}~~\exp\{\gamma_\alpha\tau+
\nonumber\\
i E_{rs}^{\alpha\pm}\tau/\hbar+
J_\alpha(\pm\tau)-\frac{ie}{\hbar}\int_{t+\tau}^t d t^\prime  g_\alpha(t^\prime) \} \Big ]
\label{eq:timeG}
\end{eqnarray}
with $\tau\leq 0$.
Since this function decays exponentially from $\tau=0$,
we now expand $P_{rr}(t+\tau)$ in Taylor series at $\tau=0$ to calculate
Eqs. (\ref{eq:master}) and (\ref{eq:tcurrent}).
By collecting the leading contributions, the occupation probabilities and tunneling currents are shown to be,
\begin{eqnarray}
\frac{d P_{ss}(t) }{dt} &=& \sum_{r\alpha,\xi=+,- } \Big [ P_{rr}(t)
 \Gamma^{\alpha \xi}_{rs}(t)
- P_{ss}(t) \Gamma^{\alpha \xi}_{sr}(t) \Big ]\nonumber\\
& &  +O({\cal H}_{SB}^4)
\label{lessP}
\end{eqnarray}
and
\begin{eqnarray}
I_\alpha^t(t) &=& e\sum_{rs} P_{rr}(t) \Big \{ \Gamma^{\alpha -}_{rs}(t)-
\Gamma^{\alpha +}_{rs}(t) \Big \}
\label{lessI}
\end{eqnarray}
where $\Gamma^{\alpha \xi}_{rs}(t) = \int_{-\infty}^0 d\tau  \Gamma^{\alpha \xi}_{rs}(t,\tau)$.
The next contributions are the fourth order of the interaction Hamiltonian ${\cal H}_{SB}(t)$
(These results can be also derived starting from the time-convolutionless solution of the density operator\cite{ahn})
and, neglecting them in the spirit of sequential tunneling, the results are now time-convolutionless. 
In the Fourier space, the time-convolutionless results read as,
\begin{eqnarray}
il\omega_A {\tilde P}_{ss}(l) &=& \sum_{r\alpha m }\sum_{\xi=+,-}\Big \{
{\tilde P}_{rr}(m) {\tilde \Gamma}^{\alpha \xi}_{rs} (l-m,0)
\nonumber\\
& & -{\tilde P}_{ss}(m) {\tilde \Gamma}^{\alpha \xi}_{sr} (l-m,0)\Big \}
\end{eqnarray}
and
\begin{eqnarray}
{\tilde I}_{\alpha}^t (l) = e \sum_{r s m }
{\tilde P}_{rr}(m) 
\Big \{ {\tilde \Gamma}^{\alpha -}_{rs} (l\!-\!m,0)\!-\!{\tilde \Gamma}^{\alpha +}_{rs} (l\!-\!m,0) \Big \}.
\end{eqnarray}

\subsection{Adiabatic limit}

As shown in Eq. (\ref{eq:timeG}),  $\Gamma^{\alpha \pm}_{rs}(t,\tau)$ is dominant around $\tau=0$
and thus,
for a slowly varying external field ($\omega_A \ll \gamma_\alpha+\pi/\hbar\beta$) it 
can be further approximated as,
\begin{eqnarray}
\Gamma^{\alpha \pm}_{rs}(t,\tau) &=&
\gamma_\alpha S_{rs}^\pm \Big [ \delta(\tau)\mp \frac{\rm csch(\pi \tau/\hbar\beta)}{\hbar\beta} 
\nonumber\\
& & {\rm Im}~~ e^{\gamma_\alpha\tau+i ( e g_\alpha(t)+E_{rs}^{\alpha\pm})\tau/\hbar+
J_\alpha(\pm\tau) } \Big ]\nonumber\\
&=&\frac{{\rm Re}}{\pi\hbar}
\int_{-\infty}^\infty d\epsilon~ \Lambda^{\alpha\pm}_{rs}(\epsilon+eg_\alpha(t)+E_{rs}^{\alpha\pm})
\nonumber\\
& & \exp \Big \{\gamma_\alpha\tau-\frac{i\epsilon}{\hbar}\tau+
J_\alpha(\pm\tau) \Big  \}
\end{eqnarray}
Namely, in this limit, energy states in the quantum dot are merely modulated by the external perturbations.
Furthermore, since these modulations are much slower than the equilibrated rate of $\gamma_\alpha+\pi/\hbar\beta$
due to tunneling and temperature, 
one can treat the problem as a static one with an additional DC-bias voltage of $ g_\alpha(t)$ at each instant. 
Then, the occupation probabilities may be determined from a static balance relation,
\begin{eqnarray}
0 = \sum_{r\alpha,\xi=+,- } \Big [ P_{rr}(t)
 \Gamma^{\alpha \xi}_{rs}(t)
- P_{ss}(t) \Gamma^{\alpha \xi}_{sr}(t) \Big ].
\end{eqnarray}

\section{ Applications of formalism }

\begin{figure}
\centering
\includegraphics[width=0.4\textwidth]{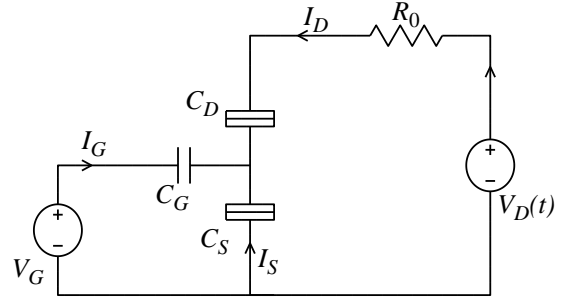}
\caption{\label{fig3}
We show a simple example of the single-electron transistor driven by a time-dependent field
via a resistor $R_0$ in the drain.
Here, symmetric tunneling barriers are assumed with their bare tunneling rates of
$\gamma_D=\gamma_S=\gamma_0/2$ and capacitances of $C_S=C_D=10^{-4}e^2/\hbar\gamma_0$.
The charging energy is chosen to be $E_C=100\hbar\gamma_0$ and
all numerical calculations are done at a temperature of $k_B T=\hbar\gamma_0$.
}
\end{figure}
As applications of our results, we now examine time-dependent currents in a single-electron transistor
based on Eqs.  (\ref{eq:master}), (\ref{eq:Gamma}), and (\ref{eq:tcurrent}).
To understand $I-V$ characteristics easily, we consider a simple circuit of Fig. 3
where a sinusoidal voltage of $v_D(t)=v_D^0 {\rm cos}(\omega_A t)$ is driven at a drain through
a resistor $R_0$ and a static voltage is applied on a gate without any impedance.
The electron-electron interaction is assumed by a Coulomb blockade model; $V_{ee}=Q^2/2C_\Sigma$.
For well-separated peaks of the Coulomb blockade oscillation, 
a charging energy ($E_C=e^2/2C_\Sigma=100\hbar\gamma_0$) is chosen to be much larger than
a broadening due to tunneling $\hbar\gamma_0$ and a thermal energy  $k_B T=\hbar\gamma_0$ considered here
($\gamma_0$ is a unit for a tunneling rate).
We also assume symmetric barriers in the drain and sources,  and denote
their tunneling rates with $\gamma_D=\gamma_S=\gamma_0/2$.
Although we assume the symmetric tunneling barriers, 
the voltage differences $g_\alpha(t)$ across each tunneling junction
as well as the characteristic impedances $Z_\alpha^t(\omega)$
 are different from each other because of a special geometry of
our circuit, so called, a common-source geometry.
In this geometry, the voltage difference of the tunnel junction to the source is mainly
determined by the gate voltage $V_G$ while the junction to the drain depends largely on
the drain voltage $V_D$.
 
\subsection{ Effects of the dissipative element $R_0$ }

As mentioned in the previous section, the first role of the dissipative elements is 
the determination of the AC voltages $g_\alpha(t)$  across each tunnel junction.
According to Eq. (\ref{eq:galpha}),
$g_\alpha(t)$ for the circuit of Fig 3 are given by, in their Fourier components,
\begin{eqnarray}
\left(
\begin{array}{c}
\tilde{g}_D(\omega)\\
\tilde{g}_S(\omega)\\
\end{array}
\right)
&=&
\left(
\begin{array}{c}
C_S+C_G\\
-C_D\\
\end{array}
\right) \frac{\tilde{v}_D(\omega)}{C_\Sigma+i w (C_G+C_S) C_D R_0}
\nonumber\\
&\sim&
\left(
\begin{array}{c}
1\\
-\frac{C_D}{C_G}\\
\end{array}
\right) \frac{\tilde{v}_D(\omega)}{1+i w C_D R_0}.
\label{role1}
\end{eqnarray}
where a relatively large capacitance of the gate capacitor compared with
those of the source and drain is assumed in the second relation.
It is noted that for a small frequency compared to a cut-off frequency $1/C_D R_0$,
the most part of $v_D$ is applied on the drain junction with a small fraction of $C_D/C_G$ 
on the source junction.
However, above the cut-off frequency, $\tilde{g_\alpha}$ fall like $1/\omega$, therefore,  a smaller
fraction of $v_D$ is applied on the tunneling junctions.

The second role of the resistance $R_0$ appears in the broadening of quasiparticle distributions
as in Eq. (\ref{eq:GammaFD}) through the probability density $P_\alpha$.
From Eq. (\ref{eq:ImZ}), $Z_\alpha^t(\omega)$ becomes for the circuit of Fig 3,
\begin{eqnarray}
\left(
\begin{array}{c}
Z_D^t(\omega)\\
Z_S^t(\omega)\\
\end{array}
\right)
=
\left(
\begin{array}{c}
1\\
\frac{C_D^2}{C_G^2}\\
\end{array}
\right) \frac{R_0}{1+i w C_D R_0}.
\label{role2}
\end{eqnarray}
The impedance $Z_S^t(\omega)$ is smaller than $Z_D^t(\omega)$ by a factor of $C_D^2/C_G^2$ and,
for $C_G\gg C_D$, we can readily set $P_S(\omega) \sim \delta(\omega)$
which implies that tunneling through the barrier to the source is irrelevant to the environment.
In Fig. 4, we show the probability density $P_D(\omega)$ at the drain and corresponding particle-distribution function
for various $R_0$ when there are no alternating perturbations.
Starting from a $\delta-$function for $R_0=0$, the probability density $P_D(\omega)$ shows 
a Lorentzian shape  in the region of positive frequencies
and exhibit exponentially decaying behavior of $P_D(-\omega)=\exp\{-\hbar\omega\beta\}P_D(\omega)$
in the negative region\cite{ingold}.
As the values of $R_0$ increases the shapes of $P_D(\omega)$ are found to become more broad
together with shifted peak positions to a positive frequency.
This means that when particles tunnel a barrier more energies should be transferred to the environment
as $R_0$ increases.
These results are also reflected in the particle(hole)-distribution functions as shown in Fig. 4-(b).
As $R_0$ increases,  the particle-distribution functions are largely depleted in the region of negative
energies.
These depletions are similar to the case as if it is a high temperature and a chemical potential is shifted to
a negative energy. Thus, one can expect that, compared to results of $R_0=0$,
tunneling currents are smeared out and start to flow at a higher drain voltage
as the resistance $R_0$ increases.
\begin{figure}
\centering
\includegraphics[width=0.4\textwidth]{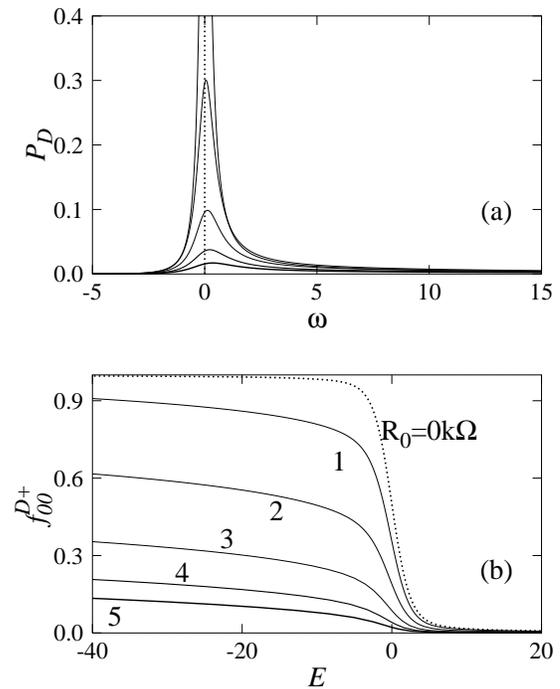}
\caption{\label{fig4}
We plot the probability density of $P_D(\omega)$ at the drain in (a) and corresponding particle-distribution
function
$f_{00}^{D+}(E)$ in (b) for $R_0=0$(dotted lines), 1, 2, 3, 4, and 5k$\Omega$(solid lines) when
there are no alternating perturbations.
}
\end{figure}

\subsection{ Tunneling currents under time-dependent fields }

As a simple example, we consider two degenerate states($\epsilon_1=\epsilon_2=0$) in the quantum dot
and, therefore, two possible many-body states with energies of $E_s= s^2 E_C (s=1,2)$ are available for tunneling.
In the static case, we obtain typical $I-V$ curves; step-like behavior and Coulomb blockade oscillations 
as a function of drain and gate voltages, respectively, for $R_0=0$ as shown in Fig. 5(dashed lines).
Under the dissipative environments($R_0 \neq 0$), the shape of the $I-V$ curves are found to be smeared out
due to the depletion of the particle(hole) distribution functions(not shown in Fig. 5).

The smearing of the tunneling currents is also found for the case of time-dependent fields.
Applying an alternating perturbation, fine structures in tunneling currents are developed because 
energy levels in the quantum dot are split into photon-side bands, $E_s+k\hbar\omega_A$ ($k=$integer).
These split energy levels are well identified in a DC part of the tunneling current
as a function of additional static drain voltages,
especially in the case of $R_0=0$(dotted line in Fig. 5-(a)).
Namely, each step appears at the voltage which gives the chemical potential of the drain corresponding to
one of energy levels, $E_s+k\hbar\omega_A$.
Under the dissipative environments, this step-like behavior is washed out as well as the height of
the steps is reduced as $R_0$ increases as shown in Fig. 5-(a)(solid lines).
Different from the static case, the reduction of the tunneling currents can be caused by the decrease of
voltage differences across tunneling barriers as well as the depletion of the particle distributions
described by Eqs. (\ref{role1}) and (\ref{role2}), respectively.
However, the results of Fig. 5 are mostly responsible for the latter effect of the environment in
this numerical simulation because the cut-off frequency of $1/R_0 C_D$ is still much larger than
the applied frequency $\omega_A$ even for $R_0=5$k$\Omega$.
Thus, we find that the tunneling currents are smeared out nearly by the same amount in both
static and time-dependent cases as $R_0$ increases.

For various values of $R_0$, we also examine a DC part of the tunneling current
as a function of a static gate voltage $V_G^0$ with a static drain voltage being zero,
and plot the results in Fig. 5-(b).
Instead of a Coulomb blockade peak in the static case(dashed line), it is found that
the tunneling currents have negative or positive values depending on the gate voltages,
and the direction is abruptly altered around the peak position.
This behavior is retained for lower frequencies of the alternating perturbation, even in adiabatic limit.
In reality, the direction of the tunneling currents is easily inferred
because the voltage difference across the tunneling junction to the source is mainly
determined by the gate voltage as mentioned in the previous section.
In the region of the negative(positive) tunneling currents, the chemical potential of the source
is lower(higher) than an energy of a relevant quantum state in the dot
and, therefore, electrons tunnel from(to) the quantum dot through the barrier connected to the source,
and vice versa.
Similar to the results as a function of a static drain voltage in Fig 5-(a), 
photon-side bands are manifested itself in steps apart from each other by $\hbar\omega_A$.
We find that these steps are well resolved in the case of $R_0=0$
while they are relatively washed out as $R_0$ increases.

For valid applications of the time-convolutionless formalism
we also calculate tunneling currents for the circuit of Fig. 3 based on 
Eqs. (\ref{lessP}) and (\ref{lessI}).
By varying parameters within small tunneling rates, it is found that the time-convolutionless formalism
give negligible differences from results obtained by the time-convolution forms.
\begin{figure}
\centering
\includegraphics[width=0.4\textwidth]{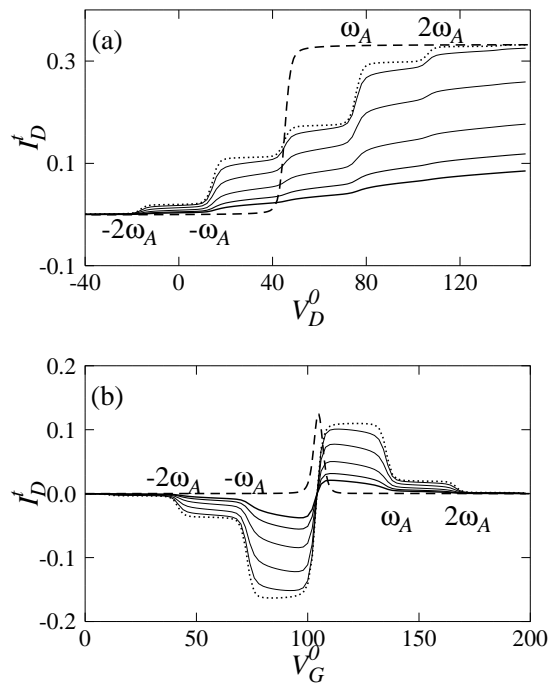}
\caption{\label{fig5}
When an alternating field is applied with $v_D^0=50 \hbar\gamma_0/e$ and $\omega_A=30\gamma_0$,
the DC part of the tunneling currents(in units of $e\gamma_0$) at the drain
are plotted as a function of a static drain voltage $V_D^0$ in (a) with $V_G^0=150\hbar\gamma_0/e$
and as a function of a static gate voltage $V_G^0$ in (b)  with $V_D^0=0$.
Here, the thicker solid line describes the result for a larger resistance of
$R_0=1, 2, 3, 4$, and $5$k$\Omega$, respectively, with the dotted line for $R_0=0$.
The step designated by $k\omega_A$ is resulted from a photon-side
band $E_1+k\hbar\omega_A$.
For a comparison, the results without the alternating field are shown in both graphs with dashed lines
for $R_0=0$ where
$V_D^0=2\hbar\gamma_0/e$ is used in (b).
}
\end{figure}

\section{Summary}

In summary, we have studied time-dependent currents of the single-electron transistor embedded in possible
dissipative circuits and driven by time-dependent perturbations.
In the regime of sequential tunneling, we present numerically tractable
forms for both displacement and tunneling currents
where the tunneling rates of Eq. (\ref{eq:Gamma}) contain explicitly 
the influence of the dissipative environments and time-dependent perturbations.
We find that the dissipative environments affect tunneling
currents of the single-electron transistor in two ways;
the determination of driving voltages at tunneling junctions
and the depletion of particle-hole distribution functions at each electrode.
Applying our formalism to a simple dissipative system and solving the problem numerically,
we show how steps in tunneling currents developed by photon-side bands are smeared out
as the system becomes more dissipative.

\acknowledgments{
This work was supported by the Korean Ministry of Science and Technology
through the Creative Research Initiatives Program under Contract No.
M1-0116-00-0008.}

\end{document}